# On the Performance of a Linearized Dual Parallel Mach-Zehnder Electro-Optic Modulator

J. Perez[1,i] and R. Llorente[2]

[1] Optical Communications Research Group, Faculty of Engineering and Environment, Northumbria University, Newcastle-upon-Tyne, NE1 8ST, UK.

[2] Nanophotonics Technology Center, Universitat Politècnica de València, C/ Camino de Vera s/n, Valencia 46022 Spain.

E-mail: joapeso@upv.es

**Abstract**—The performance of a dual parallel differential Mach-Zehnder modulator broadband linearization architecture is analysed. This study provides experimental and analytical results showing an enhancement up to 20 dB in the 3rd-order intermodulation distortion factor at 5 GHz using RF and optical asymmetrical feeding factors.

## 1. Introduction

Over the last decade, a number of bandwidth hungry applications and services have appeared, e.g., HD streaming, mobile broadband and online gaming. To address this high bandwidth demand, ubiquitous and high data-rate access technologies have emerged based on a range of wireless, fibre and optical-wireless technologies, e.g. radio-over-fibre (RoF). However, RoF systems performance, e.g. dynamic range, is limited by the nonlinear response of the optical transmitters, e.g. electro-optical modulator (EOM). EOM exhibit an inherent non-linear behaviour, which is reflected in a non-linear transfer function due to different factors, e.g., bias wavelength and polarization dependence or environmental factors [1]. Therefore, this non-linear behaviour limits the EOM dynamic range (DR) and degrades modulation performance, improving the presence of intermodulation distortion (IMD) in analog applications where EOM is a key element, such as wide-band analog fibre optic links.

Several techniques have been proposed to linearize the EOM transfer function using pre-distortion and post-distortion electronic circuit. For example, performing a dynamic bias control [2] or designing an appropriate electrode to decrease wavelength EOM dependence [3], which limits the EOM optical bandwidth. On the other hand, optical linearization techniques [1], based on parallel and serial Mach-Zehnder modulator (MZM) architectures achieve broadband EOM linearization [4-7].

## 2. Linearization technique

The aim of this letter is to analyse an optical broadband linearization technique, based on the dual parallel differential MZM (DPD-MZM) architecture shown in Figure 1. A similar dual MZM structure has been proposed previously in [5, 6] and recently on a single-drive dual-parallel MZM [7, 8]. The proposed DPD-MZM architecture considers two parallel MZM with asymmetric optical signal ($α$) and electrical radiofrequency (RF) signal ($β$) feeding. A parallel MZM architecture [9] compared with a single MZM achieves an improvement of 10 dB spurious-free dynamic range (SFDR), as shown in [3]. In the proposed scheme, the use of opposite quadrature bias point (QB) and optical and electrical asymmetric feeding in the DPD-MZM architecture results in a significant increase of the SFDR as it is stated in this paper, as SFDR is an important quality factor in analog systems [10]. The proposed linearization technique is based on the asymmetric electrical and optical feeding of each MZM in a dual parallel architecture. This technique performs a maximum optical power in one branch and minimum on the toher thanks to the factor alpha, as depicted in Figure 1.If in one MZM branch of the DPD-MZM the optical power is maximum (alpha factor) then the electrical RF signal is the lower value due to the inverse asymmetric electrical feeding factor beta. This inverse and asymmetrical behaviour allow to minimize the fifth order intermodulation harmonic components in the MZM branch with lower electrical RF signal and minimize the RF power of the third order intermodulation harmonic components on the other MZM branch with low optical signal. After the photodetection a linearized RF signal with improved rejection to the third order intermodulation harmonics will arise, as proposed in Figure 1. Moreover, the proposed DPD-MZM scheme can be used to any multioctave optical link to obtain a high dynamic-range system.

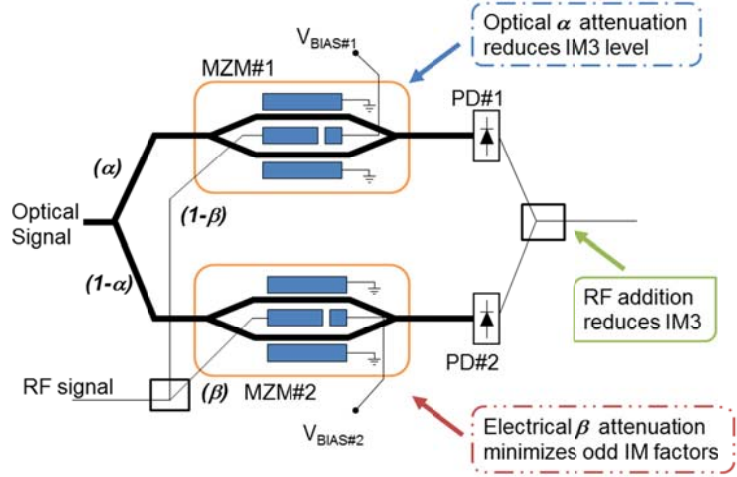

**Figure 1.** Proposed DPD-MZM internal architecture, where PD#i is photodetector i.

Let us first analyse a single MZM. In this case, the power transfer function is given by:

$$P_{out} = P_{in} \cdot L_{ff} \left( \left( 1 + \cos\left( \pi/V_{pi} (V_1 - V_2) \right) \right) \right), \tag{1}$$

where $L_{ff}$ includes modulator losses, and $\pi/V_{pi}(V_1 - V_2) \propto m_{DC} + m_1 \sin(\omega_1 t + \phi_1) + m_2 \sin(\omega_2 t + \phi_2)$ with $m_i = V_i/V_{pi}$ modulation index, and $\phi_i$ and $\omega_i$ re the phase and frequency of the RF signal on each modulator, respectively.

The expression (1) can be rewritten in terms of Bessel functions [11] as follows:

$$P_{out} = P_{in} \cdot L_{ff} \cdot \left( \begin{array}{l} 1+\cos(m_{DC}) \left( \begin{array}{l} \sum_{j=-\infty}^{\infty}\sum_{k=-\infty}^{\infty} J_{2j}(m_1) J_{2k}(m_2) \cos(2j\omega_1 t + 2k\omega_2 t) + \\ + \sum_{j=-\infty}^{\infty}\sum_{k=-\infty}^{\infty} J_{2j+1}(m_1) J_{2k+1}(m_2) \cos((2j+1)\omega_1 t + (2k+1)\omega_2 t) \end{array} \right) + \\ + \sin(m_{DC}) \left( \begin{array}{l} \sum_{j=-\infty}^{\infty}\sum_{k=-\infty}^{\infty} J_{2j+1}(m_1) J_{2k}(m_2) \sin((2j+1)\omega_1 t + 2k\omega_2 t) + \\ + \sum_{j=-\infty}^{\infty}\sum_{k=-\infty}^{\infty} J_{2j}(m_1) J_{2k+1}(m_2) \cos(2j\omega_1 t + (2k+1)\omega_2 t) \end{array} \right) \end{array} \right). \quad (2)$$

From the above expression, non-linear distortion IMD terms related to third order intermodulation distortion (IM$_3$) and third harmonic order (HD$_3$), can be easily identified. When the single MZM is biased at $m_{DC} = \pi/2$ or QB, only odd non-linear terms are included in equation (2).

Let us now evaluate the dependence of (2) with the asymmetric feeding factors at the proposed dual DP-MZM scheme. Given two parallel MZM biased at QB, and assuming that the optical signal is photodetected by an ideal photodetector with responsivity, $\Re$, equations (3) and (4) reflect 1$^{st}$- and 3$^{rd}$-order terms present in the photodetected current signal of dual PD-MZM:

$$i_{out}(\omega_1) = P_{opt} \Re L_{ff} \begin{bmatrix} -\alpha J_1(\beta m) J_0(\beta m) + \\ +(1-\alpha) J_1((1-\beta)m) J_0((1-\beta)m) \end{bmatrix}, \quad (3)$$

$$i_{out}(2\omega_1 - \omega_2) = P_{opt} \Re L_{ff} \begin{bmatrix} \alpha J_2(\beta m) J_1(\beta m) + \\ +(1-\alpha) J_2((1-\beta)m) J_1((1-\beta)m) \end{bmatrix}, \quad (4)$$

where; $\{\alpha, \beta\} \in [0,1]$. Therefore, the main 3$^{rd}$-order distortion factors can be expressed as:

$$\begin{array}{l} IM_3 = i_{out}(\omega_1)/i_{out}(2\omega_1 - \omega_2) \\ HD_3 = i_{out}(\omega_1)/i_{out}(\omega_3) \end{array}. \quad (5)$$

Simulations have been performed to determine $\alpha$ and $\beta$ factors that maximize IM$_3$ and HD$_3$ factors under ideal conditions, based on expressions (3)-(5). The IM$_3$ and carrier (H$_0$) simulated results are show in Figure 2.

In Figure 2 it can be observed than selecting a 15% electrical feeding and 65% optical feeding factors provides rejection of nearly 60 dB IM$_3$ with a 60 dB H$_0$. Moreover, considering other pair of $\alpha$ and $\beta$ feeding factors to achieve maximum IM$_3$ value would decrease H$_0$ power. That constraint has to be considered in order to analyse and implement a real DPD-MZM architecture.

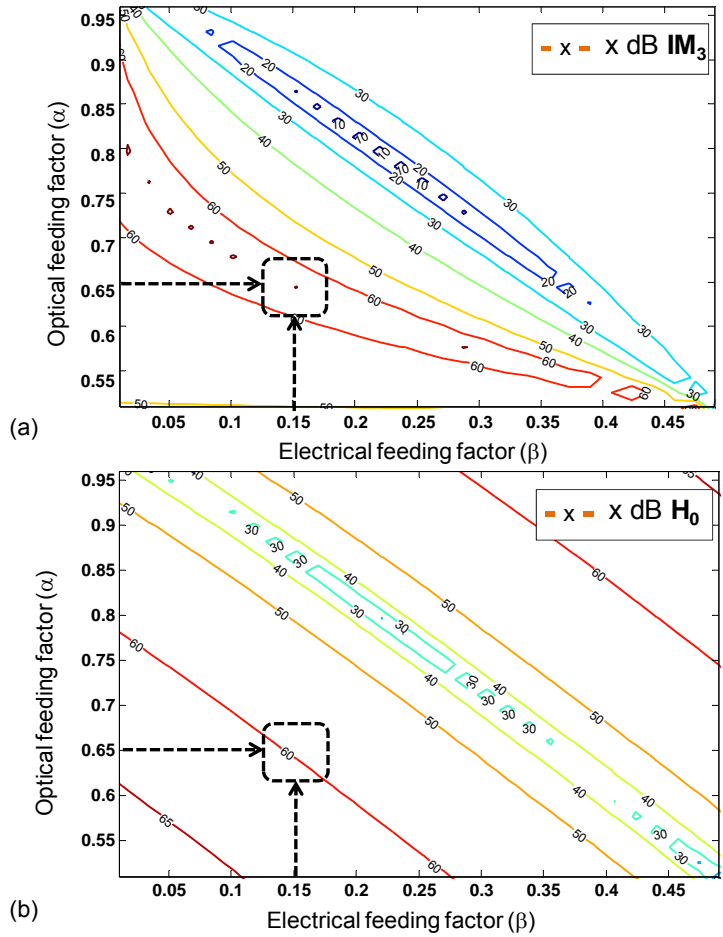

**Figure 2.** (a) IM$_3$ and (b) H$_0$ simulated values for a DPD-MZM scheme with 2 RF signals 1 GHz and 1.1 GHz and value region for IM$_3$ and H$_0$ according to selected electrical and optical feeding factors 0.65 and 0.15.

These ideal simulated results have been checked on a commercial Split-Step Fourier software simulation tool [9]. This simulation considers a dual parallel architecture with a MZM of 20 dB extinction ratio, 6 dB of insertion losses and $V_\pi$ near 1.5 V$_{DC}$. The simulation set-up comprises two ideal RF sources ($\omega_1$ and $\omega_2$) which allow us to evaluate IM$_3$ and HD$_3$. Optical attenuators and electrical delay lines are used to implement the RF and optical asymmetric feeding.

The $\alpha$ and $\beta$ asymmetrical feeding factors have been properly selected in order to compare with the ideal feeding factors simulated previously.

These non-ideal simulation results, as shown in Table 1, agree with the behaviour of ideal simulation. For a 3 dB electrical attenuation ($\beta = 0.25$) and 18.5 dB optical attenuation ($\alpha = 0.49$) values, a simulated $3^{rd}$ harmonic distortion relation of 37.63 dB is obtained, which is slightly lower than ideal simulation results. These different results can be explained by taking into account insertion losses, bias voltage deviation and other sources of noise that have been considered by the optical network simulator tool, in this case VPItransmissionMaker.

**Table 1.** Simulation results for DPD-MZM configuration for non-ideal conditions

| RF att. ($\beta$, dB) | Optical att. ($\alpha$, dB) | $HD_3$ (dB) |
|---|---|---|
| 3 | 18.5 | 37.63 |
| 4 | 20.5 | 39.60 |
| 5 | 21 | 41.66 |
| 6 | 25 | 43.67 |
| 7 | 26 | 45.60 |

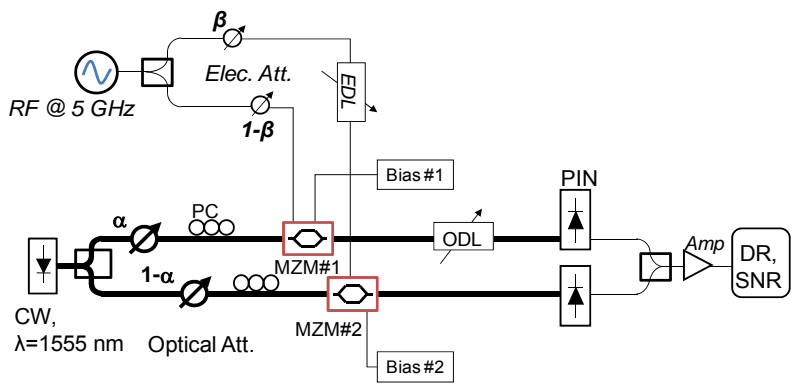

**Figure 3.** DPD-MZM experimental set up used at laboratory.

In order to validate simulation results an experimental study has been carried out. Figure 3 shows the experimental set-up for the proposed DPD-MZM scheme. This set-up comprises two commercial MZM with a low $V_\pi$ from 1.5 to 2 $V_{DC}$ and 6 to 8 dB insertion losses, both are biased at QB point to minimize $IM_2$ [1, 5]. The polarization voltage Bias#1 and Bias#2 were provided in order to accomplish with the expected QB point for each MZM around 1.75 $V_{DC}$. Moreover, this set-up includes a 6 dBm CW laser source at 1555 nm and a single-tone RF source centred at 5 GHz with a maximum output optical power of 15 dBm. The asymmetric feeding is achieved by configuring different RF and optical attenuation levels. The range of the RF attenuation is fixed at a range between 2 and 6 dB, whereas the optical attenuation is between 0.5 to 22 dB. The electrical attenuation related to the asymmetrical feeding factor was fixed by a variable step RF attenuator and the optical attenuation was performed by a variable single mode inline fibre optical attenuator, as depicted in Figure 4. The used of an electrical delay line (EDL) and an optical delay line (ODL) of 330 ps has been considered in this experimental set-up in order to adjust and minimize the effect of the temporal drifts in the optical and RF electrical signals due to different lengths of the paths for each parallel MZM branch. Figure 4 shows $HD_3$ results for different experimental values of RF and optical attenuation. The 6 dB RF fixed attenuation simulation results are also included. It can be observed that the experimental DPD-MZM architecture obtains a $HD_3$ improvement close to 20 dB over $HD_3$ single MZM result that was 31.83 dB at $HD_3$, as depicted in Figure 4.

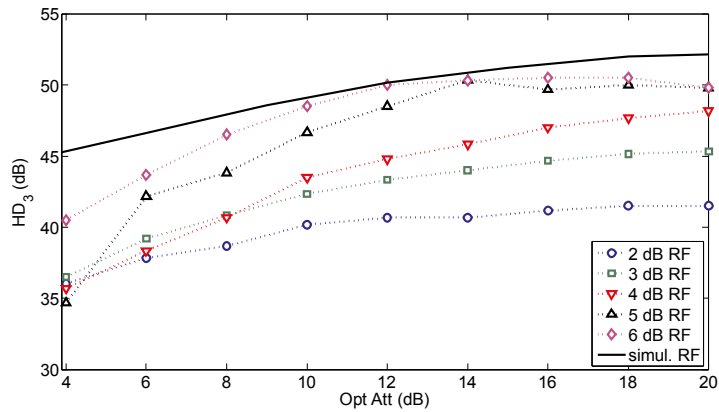

**Figure 4.** Experimental (dotted line) and simulated $HD_3$ results versus different optical attenuation values for a fixed RF attenuation.

Experimental results show that asymmetric feeding values should move between $\alpha = [0.66-0.95] \rightarrow [16-21]$ dB and $\beta = [0.05-0.20] \rightarrow [5-8]$ dB in order to maximize $HD_3$. These experimental asymmetric factors agree with the simulated results. Moreover, the PD-MZM performance can be enhanced carrying a balanced detection [12] and implementing a push-pull configuration on parallel MZM architectures to improve dynamic range [13].

Figure 5 shows experimental and simulation carrier power ($H_0$) results. The experimental values of $H_0$ are slightly higher than the simulated one due to an overestimation of the optical and electrical sources in the simulation tests. In this figure, $H_0$ decreases with asymmetric feeding factors that maximize $HD_3$. This constraint has to be considered in order to face the design of an integrated DPD-MZM architecture in one device.

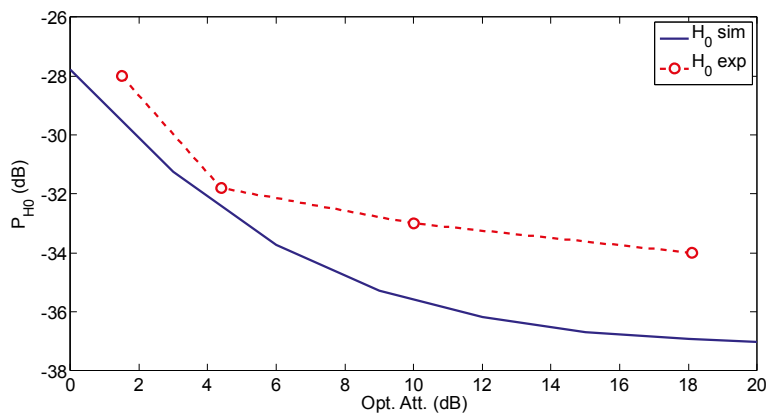

**Figure 5**. Simulation and experimental carrier power level $H_0$ with 6 dB fixed RF attenuation

The experimental results indicate that this technique improves the third harmonics rejection on a microwave carrier with improvements in the range of 15-20 dB compared with a single MZM. On the other hand it suppose a first step on the implementation of dual parallel linearization technique on a more robust design that includes components off-the-self. On the other hand, next steps will drive to an integration of this system on a one device dual-parallel MZM that will include all the considerations of differential photodetection and others in order to improve the behaviour of the linearization technique in broadband communications systems.

## 3. Conclusions

The proposed dual configuration achieves a 20 dB $HD_3$ improvement over a single QB biased MZM modulator, when using properly designed RF and optical asymmetrical feeding factors. Simulated and experimental results show that this DPD-MZM architecture is an efficient broadband linear technique to improve MZM modulator performance. The results of the analytical and simulation evaluation demonstrated that this architecture is valid for broadband communications systems and the experimental set-up indicates the feasibility to implement this linearization technique with a microwave carrier of 5 GHz.

## Acknowledgements

Support by Spanish MINECO Juan de la Cierva fellowship JCI-2012-14805 and Spanish National Plan project MODAL TEC2012- 38558-C02-01 are acknowledged.

---

[i] *J. Perez is now with the Optical and Quantum Communications Group, ITEAM Research Institute, Universitat Politècnica de València, C/ Camino de Vera, s/n, 46022, Valencia, Spain, email: **joapeso@upv.es***